\documentclass[3p,times,twocolumn]{elsarticle}

\usepackage{ecrc}

\volume{00}

\firstpage{1}

\journalname{Nuclear Physics B Proceedings Supplement}

\runauth{}

\jid{nuphbp}

\jnltitlelogo{Nuclear Physics B Proceedings Supplement}

\usepackage{amssymb}

\usepackage[figuresright]{rotating}

\begin{document}

\begin{frontmatter}



\dochead{}

\title{Stability of black holes and solitons in Anti-de Sitter space-time }


\author{Betti Hartmann}

\address{School of Engineering and Science, Jacobs University Bremen, 28759 Bremen, Germany}

\begin{abstract}
The stability of black holes and solitons in 
d-dimensional Anti-de Sitter (AdS$_d$) space-time against scalar field condensation is discussed.
The resulting solutions are ``hairy'' black holes and solitons, respectively.
In particular, we will discuss static black hole solutions with hyperbolic, flat and spherical horizon topology and
emphasize that two different type of instabilities exist depending on whether the
scalar field is charged or uncharged, respectively. We will also discuss the influence of Gauss-Bonnet curvature 
terms. The results have applications within the AdS/CFT correspondence
and describe e.g. holographic insulator/conductor/superconductor phase transitions.

\end{abstract}


\end{frontmatter}


\section{Introduction}
Black holes are very intriguing objects that constitute exact solutions of Einstein's equations
(and generalizations thereof). It is well known that 4-dimensional black holes of the Einstein-Maxwell
equations are uniquely characterized by a small number of quantities that are subject to a Gauss' law. 
These are the mass $M$, the charge $Q$ and the angular momentum $J$. Furthermore, black holes
possess a well defined temperature $T=\hbar \kappa/(2\pi k_B c)$ and entropy $S=k_B c^3 A/(4 G \hbar)$
given in terms of the surface gravity $\kappa$ and the area of the black hole horizon $A$. Here $\hbar$ is Planck's constant,
$k_B$ Boltzmann's constant, $c$ the speed of light in vacuum and $G$ Newton's constant, respectively. While
the appearance of $\hbar$, $G$ and $c$ in these relations indicates clearly that quantum physics and 
General Relativity are involved, the appearance of $k_B$ is more surprising since this is usually connected
to thermodynamical system. Studying the laws of Black hole mechanics more closely and comparing them
to the laws of thermodynamics reveals that their is an apparent resemblance. However, it is important to note 
that the entropy of a black hole is encoded in a 2-dimensional surface $A$ - very unlike the entropy of
a standard thermodynamical system which is proportional to the 3-dimensional volume. This already hints to
the fact that black holes are special systems that show ``holographic'' features.

Since the advent of the gravity/gauge theory duality \cite{ggdual} and in particular
the Anti-de Sitter/Conformal Field Theory (AdS/CFT) correspondence \cite{adscft} it is now obvious 
that holographic dualities
play a very important r\^ole in black hole physics. The AdS/CFT correspondence states that a gravity theory
in a d-dimensional asymptotically AdS space-time is dual to a conformal quantum field theory on the (d-1)-dimensional
boundary of AdS. This duality is a weak-strong coupling duality
which allows to study strongly coupled Quantum Field Theories with the help
of black hole and soliton solutions of standard General Relativity. This has e.g. been used to
study the phenomenon of high-temperature superconductivity \cite{gubser,hhh,reviews}.
In these models the black hole becomes unstable to form non-trivial fields outside its
horizon when being close to extremality. The value of these bulk fields on the AdS boundary
then correspond to sources in the dual theory. As such it is of importance to understand the stability
of black holes and solitons in AdS space-time. A central r\^ole in the discussion
of stability of AdS black holes with respect to formation of scalar fields
outside the horizon plays the Breitenlohner--Freedman (BF) bound \cite{bf}. This states that for scalar
mass $m^2 \ge m^2_{\rm BF,d} = -(d-1)^2/(4L^2)$ the d-dimensional AdS space-time is stable. Here $L^2$ is
the AdS curvature radius. In fact, two different type of instabilities appear for AdS black holes depending
on whether the scalar field is uncharged or charged, respectively. For an 
{\it uncharged} scalar field the instability is related to the fact that close to extremality the
near-horizon geometry is typically given by AdS$_2\times M_{d-2}$, where $M_{d-2}$ is a (d-2)-dimensional
manifold \cite{Robinson:1959ev,Bertotti:1959pf,Bardeen:1999px}, while the black hole is asymptotically AdS$_d$. Now, choosing the scalar mass $m^2$ above the
d-dimensional BF bound, but below the 2-dimensional BF bound will make the black hole become
unstable to formation of scalar hair close to the horizon, while it still asymptotes to AdS. 
In \cite{gubser} it was realized that an AdS black hole can also become unstable to the formation
of a {\it charged} scalar field. This is connected to the coupling between the gauge field and the
scalar field which leads to an effective mass term $m^2_{\rm eff}=m^2 - e^2 \vert g^{tt}\vert A_t^2$
in the equation for the scalar field. Here $e^2$ is the gauge coupling, $A_t$ the time-component of the
gauge potential and $g^{tt}$ the $tt$-component of the metric tensor. Now, close to the horizon
of a black hole this component can become very large such that $m^2_{\rm eff}$ drops below the
BF bound, while asymptotically $m_{\rm eff}^2=m^2 \ge m_{\rm BF,d}^2$. Here, I will discuss both type of instabilities for different type of black holes and solitons
in AdS space-time. 
Another interesting question related to the application within the AdS/CFT correspondence is how the obtained results
depend on the gravity model used. Within the gauge/gravity duality the couplings are related as follows
$\lambda=(L/l_s)^4=g^2 N$, $g_s\sim g^2$, where $\lambda$ is the 't Hooft coupling, $N$ the rank of the 
gauge group with gauge coupling $g$, $l_s$ the string scale and $g_s$ the string coupling. 
Now, the limit $g_s\rightarrow 0$, $l_s/L \rightarrow 0$ corresponds to standard Einstein gravity which
is dual to a strongly coupled Quantum Field Theory with $N\rightarrow \infty$. Taking $1/N$ corrections into  
account would then correspond to considering higher order curvature corrections on the gravity side. 
Here, I will discuss the influence of the Gauss-Bonnet term that appears as first order correction in
certain String Theory models.  

In Section 2, I will give the model, while I will discuss recently obtained results in Section 3.
I conclude in Section 4.  

\section{The Model}
The model that was studied extensively in 
\cite{Brihaye:2012ww,Brihaye:2012cb,Brihaye:2011fj,Brihaye:2011hm,Brihaye:2010mr} is given by the following
action:
\begin{eqnarray}
S &=&  \int d^5 x \sqrt{-g} \left(16\pi G {\cal L}_{\rm m} + R - 2 \Lambda \right. \nonumber \\
&+& \left. 
\frac{\alpha}{4}\left(R^{MNKL} R_{MNKL} - 4 R^{MN} R_{MN} + R^2\right)\right)
\end{eqnarray}
with $M,N,K,L=0,1,2,3,4$ and matter Lagrangian
\begin{equation}
{\cal L}_{\rm m}= -\frac{1}{4} F_{MN} F^{MN} - \left(D_M\psi\right)^* D^M \psi - m^2 \psi^*\psi  \ ,
\end{equation}
where $F_{MN} =\partial_M A_N - \partial_N A_M$ is the U(1) field strength tensor, 
$D_M\psi=\partial_M \psi - ie A_M \psi$ is the covariant derivative,
$\Lambda=-6/L^2$ is the cosmological constant, 
$G$ Newton's constant, $\alpha$ the Gauss--Bonnet coupling, 
$e$ the gauge coupling and
$m^2$ the mass of the scalar field $\psi$. We have added a Gauss-Bonnet term in order to be able to
construct solutions away from ``pure'' General Relativity which corresponds to studying $1/N$ corrections in the
dual theory. The motivation within the application of holographic superconductors is the Coleman-Mermin-Wagner
theorem \cite{CMW} which states that spontaneous symmetry breaking of a continuous symmetry is forbidden
in (2+1) dimensions at finite temperature. In contrast to that holographic superconductors as duals
of black holes in Einstein gravity have been constructed \cite{hhh}. The question is then
whether higher order curvature corrections can suppress condensation. This has been addressed in the
``probe limit'' in \cite{Gregory:2009fj}. Here, we are interested in the full back-reacted case.
 
We are considering static solutions and hence choose the following Ansatz for the
metric
\begin{equation}
ds^2 = - f(r) a^2(r) dt^2 + \frac{1}{f(r)} dr^2 + \frac{r^2}{L^2} d\Sigma^2_{k,3}
\end{equation}
with
$d\Sigma^2_{-1,3}=d \Xi^2_{3}$ for the hyperbolic case ($k=-1$), 
$d\Sigma^2_{0,3}=dx^2 + dy^2 + dz^2$ for the flat case ($k=0$) and
$d\Sigma^2_{1,3}=d \Omega^2_{3}$ for the spherical case ($k=1$). 
For the matter fields we have 
\begin{equation}
 A_M dx^M = \phi(r) dt \ \ , \ \  \psi=\psi(r) \ .
\end{equation} 
Note that $\psi(r)$ can be chosen to be real due to U(1) gauge invariance of the model. 
The resulting
system of ordinary differential equations has to be solved numerically subject to appropriate boundary 
conditions. On the AdS boundary the matter functions have the following behaviour
\begin{equation}
  \phi(r\gg 1) = \mu - \frac{Q}{r^2}  \ \ , \ \ 
  \psi(r\gg 1) = \frac{\psi_{-}}{r^{\lambda_{-}}} + \frac{\psi_{+}}{r^{\lambda_{+}}} \ \
\end{equation}
where $\lambda_{\pm} = 2 \pm \sqrt{4 + m^2 L_{\rm eff}^2}$
with effective AdS radius $L_{\rm eff}^2 \equiv \frac{2 \alpha}{1 - \sqrt{1 - 4 \alpha/L^2}} 
       \sim L^2 \left(1  -  \alpha/ L^2 + O(\alpha^2)\right)$. $\mu$ corresponds to the
chemical potential and $Q$ to the charge for $k=1$ and to the charge density for $k=-1,0$, respectively.
The metric functions fulfill
 \begin{eqnarray}
 f(r\gg1)&=& k +\frac{r^2}{L_{\rm eff}^2} + \frac{f_2}{r^2} + O(r^{-4}) \\ 
a(r\gg 1)&=&1+\frac{a_4}{r^4} + O(r^{-6}) \ ,
\end{eqnarray}
where $f_2$, $a_4$ are constants that have to be determined numerically. 
The energy $E$ per unit three volume $V_3$ can be given in terms of these constants and reads 
\begin{equation}
E/V_3=\sqrt{1-\alpha/L^2}(-3f_2 - 8 a_4/L_{\rm eff}^2).
\end{equation}
The free energy then is $F=E-TS-\mu Q$, where
\begin{equation}
 T=\frac{f'(r_h) a(r_h)}{4\pi} \ \ , \ \ \frac{S}{V_3}=\frac{r_h^3}{4G}(1-6\alpha/r_h^2)
\end{equation}
are the temperature and entropy of a
black hole with horizon radius $r_h$, respectively. 

For $m^2 > m_{\rm BF,2}^2$ the black holes do not form scalar hair and hence $\psi(r)\equiv 0$. In this
case an explicit solution of the coupled equations exists. This is given by 
\cite{deser,Wheeler:1985nh,wiltshire,Cai:2001dz}
\begin{eqnarray}
\phi(r)&=&\frac{Q}{r_h^2}-\frac{Q}{r^2} \ \ , \ \ a(r)\equiv 1 \nonumber \\
f(r)&=&k + \\ \nonumber 
&+&\frac{r^2}{2\alpha}\left(1-\sqrt{1-\frac{4\alpha}{L^2} + \frac{4\alpha M}{r^4} - \frac{64\pi\alpha G Q^2}{r^6}}\right) \ .
\end{eqnarray} 
In the following, we will discuss black hole and soliton solutions with scalar hair which can only be obtained
numerically. The limit $G=0$ (which corresponds to $e=\infty$ due to scaling symmetries present in the model) 
is often called the ``probe limit'' in the context of holographic applications. In this case, the 
space-time is fixed and does not back-react on the matter fields. But even in this case,
the resulting coupled scalar-gauge field equations have to be solved numerically. 

\subsection{Hyperbolic case, $k=-1$}
Here we consider uncharged black holes ($Q=0$) and an uncharged scalar field ($e=0$).
The uncharged $k=-1$ case is interesting since (other than the $k=0,1$ cases) it possesses an extremal limit.
It has been considered in \cite{Dias:2010ma} as a toy-model for uncharged rotating black holes.
The extremal solution has $T=0$ and $r_h^{(ex)}=L/\sqrt{2}$. Close to extremality the near-horizon
topology is AdS$_2 \times H^3$ \cite{aste}. In \cite{Brihaye:2011hm} we studied hyperbolic Gauss-Bonnet 
black holes. We found that the near-horizon geometry in the extremal limit contains an AdS$_2$ factor, where
the curvature radius is given by $R=\sqrt{L^2/4-\alpha}$. For $m^2$ larger than $m_{\rm BF,5}^2=-4/L_{\rm eff}^2$ 
but smaller than $m_{\rm BF,2}^2=-1/(4R^2)$ we would thus expect the black holes to become unstable with respect to
scalar hair formation at the horizon. That is what we have shown numerically in \cite{Brihaye:2011hm}. 
We find that the black holes with scalar hair are thermodynamically preferred since their free energy
is {\it always} lower than that of the black holes without scalar hair. We also observed that
the larger the Gauss-Bonnet coupling $\alpha$ the lower is the temperature $T$ at which the instability appears.

\subsection{Planar case, $k=0$}
The planar case has applications to high-temperature superconductivity (see e.g. \cite{hhh} and references therein).
The reason for this is that on the boundary of AdS with $r \rightarrow \infty$ the strongly coupled
dual theory ``lives'' on a plane and high-temperature superconductivity is mainly associated to 2-dimensional
layers within the superconducting material, e.g. to the copper-oxide planes in a cuprate superconductor. 
As such the formation of scalar hair on a charged AdS black hole
has been associated to a conductor/superconductor phase transition. Interestingly, this model also
allows soliton solutions. These are obtained by double Wick-rotating the Schwarzschild-AdS solution. This
solution has a periodic coordinate which is necessary to avoid a conical singularity. Since solitons
do not have a temperature associated to them, this model can be considered at any $T$, in particular at $T=0$.
As shown in \cite{witten2} this soliton has a strictly positive and discrete spectrum and as such possesses
an energy gap. However, the soliton becomes unstable to the formation of scalar hair if the chemical potential is large enough 
\cite{Nishioka:2009zj,Horowitz:2010jq,Brihaye:2011vk}. This has been interpreted as a holographic insulator/superconductor
phase transition. Note that these transitions are in particular possible at $T=0$ which corresponds
to a quantum phase transition that is driven by quantum - not thermal - fluctuations.
Comparing the soliton and black hole solutions with and without scalar hair with respect
to their free energy (in the canonical ensemble) it is then possible to decide which phase
dominates at a given temperature and chemical potential \cite{Horowitz:2010jq,Brihaye:2011vk}. The resulting
phase diagrams possess two triple points at small values of the back-reaction. However, at large back-reaction
one phase disappears completely from the phase diagram and the qualitative features resemble closely those
of the phase diagrams of high-temperature superconductors. Furthermore, new type of phase transitions can be predicted.

Even in the back-reacted case and in the presence of the Gauss-Bonnet term 
black holes form scalar hair for $T > 0$. In the AdS/CFT language this means that the condensation
gets harder away from the probe and the $N\rightarrow \infty$ limit, respectively, but that condensation 
cannot be suppressed \cite{Brihaye:2010mr}.

\subsection{Spherical case, $k=1$}
Here we discuss solutions in asymptotic global AdS$_d$. 
As was shown in \cite{Basu:2008st,Dias:2011tj,Brihaye:2012cb} soliton solutions with scalar hair
can be constructed in this case. However, these exist only on a limited domain of the
$Q$-$e^2$-plane. At a critical value of $e$ which depends on both $Q$ as well as on the Gauss-Bonnet 
coupling $\alpha$ the numerical
results suggest that the soliton with scalar hair reaches a singular solution with $a(0)\rightarrow 0$. Furthermore,
it was found \cite{Brihaye:2012cb} that the range of allowed $Q$ and $e$ values is enlarged when including
a Gauss-Bonnet term. 

Black hole solutions with scalar hair also exist in this case. For $\alpha=0$ these tend to the soliton
solutions in the limit $r_h\rightarrow 0$, while for $\alpha > 0$ the behaviour in this limit depends
strongly on the size of the Gauss-Bonnet coupling. For small $\alpha$ solutions exist all the way down
to $r_h\rightarrow 0$ suggesting that they tend to the corresponding Gauss-Bonnet soliton. However, the results
in \cite{Brihaye:2012cb} suggest that while a corresponding Gauss-Bonnet soliton exists for the given
parameters, the black hole does {\it not} tend to this solution in the limit $r_h\rightarrow 0$. 
For $\alpha$ large, on the other hand, the black hole solutions tend to a solution with $a(r_h)\rightarrow 0$
at a critical and non-vanishing value of the black hole horizon radius $r_h$. However, this limiting
solution can {\it not} be an extremal Gauss-Bonnet black hole with scalar hair. It was shown in \cite{Brihaye:2012cb}
that extremal Gauss-Bonnet black holes do {\it not} support scalar hair in their near-horizon geometry. 

\subsection{Connection between the $k=0$ and the $k=1$ case}
It was realized in \cite{Gentle:2011kv} for $\alpha=0$ and in \cite{Brihaye:2012cb} for $\alpha\neq 0$ that
the $k=1$ and the $k=0$ AdS black holes are connected to each other. Applying a rescaling $r\rightarrow \lambda r$, $t\rightarrow \lambda^{-1}t$,
$M\rightarrow \lambda^4 M$, $Q\rightarrow \lambda^3 Q$ for a $k=1$ solution and letting $\lambda \rightarrow \infty$
we recover the $k=0$ solution. This limit was hence called ``the planar limit'' \cite{Gentle:2011kv}.
For $\lambda\rightarrow \infty$ we obviously have $Q\rightarrow \infty$ and the solution becomes comparable in 
size to the AdS radius of the space-time. This is possible because the
electromagnetic repulsion is balanced by the gravitational attraction present in AdS. 

\section{Conclusion}
As shown in this talk there are two mechanism that can make AdS black holes unstable to scalar hair formation.
An uncharged scalar field can form on a near-extremal black hole with an AdS$_2$ factor in its near horizon geometry.
The example discussed here is an uncharged, static black hole with hyperbolic horizon. This seems to be a generic
feature of near-extremal black holes and is fundamentally connected to the fact that the BF bound decreases
with increasing space-time dimension $d$. This applies also to other type of black hole solutions that have a
near-horizon geometry containing an AdS$_2$ factor. It would be interesting to see how these results
extend to rotating and even time-dependent black holes. First steps in this respect have been
done in \cite{Dias:2010ma,sonner,Brihaye:2012ww}, but a lot of work is still necessary. 

A black hole
becomes unstable to the formation of a charged 
scalar field on its horizon due to the coupling of the scalar field to a gauge field and hence
the appearance of an $r$-dependent effective mass term. The examples discuss here are black holes
with planar and spherical horizon topology. 
Remarkably, the black holes with scalar hair are thermodynamically preferred.  

Black holes and their corresponding solitonic counterparts in planar AdS have been used to 
describe condensed matter phenomena holographically. In this approach the scalar field has been considered
to be an order parameter that vanishes above a critical temperature for black holes and below a
critical chemical potential for solitons. In the case of the charged scalar field the $e=0$ limit is 
special. In this limit, the scalar field must be chosen complex (due to the global U(1) symmetry present in the model).
Typically, the scalar field is parametrized with a time-dependent phase and the resulting solutions are 
AdS boson stars \cite{radu,hartmann_riedel,hartmann_riedel2,Brihaye:2013hx}. 
These solutions have recently played an important r\^ole in the study of the stability of the AdS
space-time. In \cite{Bizon:2011gg} evidence for instabilities of asymptotically AdS to the formation
of black holes was claimed. Following this investigation, the stability of boson stars in AdS was studied 
(see e.g.
\cite{Dias:2012tq,Buchel:2012uh,Buchel:2013uba}) and it was shown that AdS boson stars are 
non-pertubatively stable \cite{Dias:2012tq}.

\section{Acknowledgments} I would first of all like to thank the organizers of the Light Cone conference
that took place in Delhi, India in December 2012 for their invitation and hospitality.
I enjoyed my stay thoroughly and hope to come visit again very soon!
Furthermore, I would like to thank Yves Brihaye and Sardor Tojiev for collaboration on this topic. 
Finally, I gratefully acknowledge the Deutsche Forschungsgemeinschaft (DFG) for financial support
within the framework of the DFG Research Training group 1620 {\it Models of gravity}.






\begin{thebibliography}{00}

\bibitem{ggdual} 
{\it see e.g.} O.~Aharony, S.~S.~Gubser, J.~M.~Maldacena, H.~Ooguri and Y.~Oz,
  Phys.\ Rept.\  {\bf 323} (2000) 183
  [arXiv:hep-th/9905111];
 E.~D'Hoker and D.~Z.~Freedman,
  arXiv:hep-th/0201253;
M. Benna and I. Klebanov, {\it Gauge-string duality and some applications} [arXiv: 0803.1315 [hep-th]].   
\bibitem{adscft} J. Maldacena, Adv. Theo. Math. Phys. {\bf 2} (1998) 231;
Int. J. Theor. Phys. {\bf 38} (1999) 1113 [arXiv:hep-th/9711200].
\bibitem{gubser} S.~S.~Gubser,
  Phys.\ Rev.\  D {\bf 78} (2008) 065034
  [arXiv:0801.2977 [hep-th]].
 \bibitem{hhh} 
 S.~A.~Hartnoll, C.~P.~Herzog and G.~T.~Horowitz, Phys. Rev. Lett. {\bf 101} (2008) 031601 [arXiv:0803.3295 [hep-th]];
  JHEP {\bf 0812} (2008) 015
  [arXiv:0810.1563 [hep-th]];
  G.~T.~Horowitz and M.~M.~Roberts,
  Phys.\ Rev.\  D {\bf 78} (2008) 126008
  [arXiv:0810.1077 [hep-th]].

\bibitem{reviews} {\it for recent reviews see} C.~P.~Herzog,
  J.\ Phys.\ A  {\bf 42} (2009) 343001;
  S.~A.~Hartnoll,
  Class.\ Quant.\ Grav.\  {\bf 26} (2009) 224002
  [arXiv:0903.3246 [hep-th]];
G. Horowitz, {\it Introduction to holographic superconductors},
arXiv:1002.1722.
 \bibitem{bf} P.~Breitenlohner and D.~Z.~Freedman,
  Annals Phys.\  {\bf 144} (1982) 249.

\bibitem{Robinson:1959ev}
  I.~Robinson,
  Bull.\ Acad.\ Pol.\ Sci.\ Ser.\ Sci.\ Math.\ Astron.\ Phys.\  {\bf 7} (1959)
351.

\bibitem{Bertotti:1959pf}
  B.~Bertotti,
  Phys.\ Rev.\  {\bf 116} (1959) 1331.




\bibitem{Bardeen:1999px}
  J.~M.~Bardeen, G.~T.~Horowitz,
  Phys.\ Rev.\   {\bf D60 } (1999)  104030,
  [arXiv:hep-th/9905099].


\bibitem{Brihaye:2012ww}
  Y.~Brihaye, B.~Hartmann and S.~Tojiev,
  Phys.\ Rev.\ D {\bf 87} (2013) 024040
  [arXiv:1210.2268 [gr-qc]].

\bibitem{Brihaye:2012cb}
  Y.~Brihaye and B.~Hartmann,
  Phys.\ Rev.\ D {\bf 85} (2012) 124024
  [arXiv:1203.3109 [gr-qc]].

\bibitem{Brihaye:2011fj}
  Y.~Brihaye and B.~Hartmann,
  JHEP {\bf 1203} (2012) 050
  [arXiv:1112.6315 [hep-th]].

\bibitem{Brihaye:2011hm}
  Y.~Brihaye and B.~Hartmann,
  Phys.\ Rev.\ D {\bf 84} (2011) 084008
  [arXiv:1107.3384 [gr-qc]].


\bibitem{Brihaye:2010mr}
  Y.~Brihaye and B.~Hartmann,
  Phys.\ Rev.\ D {\bf 81} (2010) 126008
  [arXiv:1003.5130 [hep-th]].

\bibitem{CMW} N.~D.~Mermin, H.~Wagner, 
Phys. Rev. Lett. {\bf 17} (1966) 1133;
S.~Coleman,
Commun. Math. Phys. {\bf 31} (1973) 259.

\bibitem{Gregory:2009fj}
  R.~Gregory, S.~Kanno, J.~Soda,
  JHEP {\bf 0910 } (2009)  010,
  [arXiv:0907.3203 [hep-th]].



\bibitem{deser}
  D.~G.~Boulware, S.~Deser,
  Phys.\ Rev.\ Lett.\  {\bf 55 } (1985)  2656.

\bibitem{Wheeler:1985nh}
  J.~T.~Wheeler,
  Nucl.\ Phys.\  {\bf B268 } (1986)  737.
  


\bibitem{wiltshire}
  D.~L.~Wiltshire,
  Phys.\ Rev.\  {\bf D38 } (1988)  2445.
  
\bibitem{Cai:2001dz}
  R.~-G.~Cai,
  Phys.\ Rev.\   {\bf D65} (2002)  084014,
  [arXiv:hep-th/0109133].




\bibitem{Dias:2010ma}
  O.~J.~C.~Dias, R.~Monteiro, H.~S.~Reall, J.~E.~Santos,
  JHEP {\bf 1011 } (2010)  036,
  [arXiv:1007.3745 [hep-th]].

\bibitem{aste} 
  D.~Astefanesei, N.~Banerjee, S.~Dutta,
  JHEP {\bf 0811 } (2008)  070,
  [arXiv:0806.1334 [hep-th]].

\bibitem{witten2}
  E.~Witten,
  Adv.\ Theor.\ Math.\ Phys.\  {\bf 2} (1998) 505
  [hep-th/9803131].

\bibitem{Nishioka:2009zj}
  T.~Nishioka, S.~Ryu and T.~Takayanagi,
  JHEP {\bf 1003} (2010) 131
  [arXiv:0911.0962 [hep-th]].


\bibitem{Horowitz:2010jq}
  G.~T.~Horowitz and B.~Way,
  JHEP {\bf 1011} (2010) 011
  [arXiv:1007.3714 [hep-th]].

\bibitem{Brihaye:2011vk}
  Y.~Brihaye and B.~Hartmann,
  Phys.\ Rev.\ D {\bf 83} (2011) 126008
  [arXiv:1101.5708 [hep-th]].

\bibitem{Basu:2008st}
  P.~Basu, A.~Mukherjee and H.~-H.~Shieh,
  Phys.\ Rev.\ D {\bf 79} (2009) 045010
  [arXiv:0809.4494 [hep-th]].

\bibitem{Dias:2011tj}
  O.~J.~C.~Dias, P.~Figueras, S.~Minwalla, P.~Mitra, R.~Monteiro and J.~E.~Santos,
  JHEP {\bf 1208} (2012) 117
  [arXiv:1112.4447 [hep-th]].

\bibitem{Gentle:2011kv}
  S.~A.~Gentle, M.~Rangamani and B.~Withers,
  JHEP {\bf 1205} (2012) 106
  [arXiv:1112.3979 [hep-th]].

\bibitem{sonner}
  J.~Sonner,
  Phys.\ Rev.\  D {\bf 80} (2009) 084031
  [arXiv:0903.0627 [hep-th]].

\bibitem{radu}  D.~Astefanesei and E.~Radu,
  Nucl.\ Phys.\ B {\bf 665} (2003) 594
  [gr-qc/0309131].
\bibitem{hartmann_riedel} 
 B.~Hartmann and J.~Riedel,
  Phys.\ Rev.\ D {\bf 86} (2012) 104008,
[arXiv:1204.6239 [hep-th]].


\bibitem{hartmann_riedel2} 
 B.~Hartmann and J.~Riedel,
  Phys.\ Rev.\ D {\bf 87} (2013) 4,  044003
  [arXiv:1210.0096 [hep-th]].


\bibitem{Brihaye:2013hx}
  Y.~Brihaye, B.~Hartmann and S.~Tojiev,
  Class.\ Quant.\ Grav.\  {\bf 30} (2013) 115009
  [arXiv:1301.2452 [hep-th]].




 \bibitem{Bizon:2011gg}
  P.~Bizon and A.~Rostworowski,
  Phys.\ Rev.\ Lett.\  {\bf 107} (2011) 031102
  [arXiv:1104.3702 [gr-qc]].
 
 
 
 
  \bibitem{Dias:2012tq}
  O.~J.~C.~Dias, G.~T.~Horowitz, D.~Marolf and J.~E.~Santos,
  Class.\ Quant.\ Grav.\  {\bf 29} (2012) 235019
  [arXiv:1208.5772 [gr-qc]].
  
  
 

\bibitem{Buchel:2012uh}
  A.~Buchel, L.~Lehner and S.~L.~Liebling,
  Phys.\ Rev.\ D {\bf 86} (2012) 123011
  [arXiv:1210.0890 [gr-qc]].

   \bibitem{Buchel:2013uba}
  A.~Buchel, S.~L.~Liebling and L.~Lehner,
  arXiv:1304.4166 [gr-qc].
  


\end{thebibliography}
\end{document}